\documentclass[12pt]{article}
\usepackage[utf8]{inputenc}
\usepackage[T1]{fontenc}
\usepackage[english]{babel}
\usepackage{amsmath,amssymb}

\usepackage{graphicx}
\usepackage{fancyhdr}
\usepackage{url}
\usepackage{cancel}
\usepackage{xcolor}
\usepackage{cite}
\usepackage{ulem}

\begin{document}
\title{Moir\'e patterns generated by stacked 2D lattices: a general algorithm to identify primitive coincidence cells}

\author{V. Carnevali$^{1,\dag}$, S. Marcantoni$^{1,2,\ddag}$, M. Peressi$^{1}$}

\maketitle
\thispagestyle{fancy}

1. Department of Physics, University of Trieste, via A. Valerio 2, 34127 Trieste, Italy\\

2. National Institute for Nuclear Physics (INFN), Trieste Section, via A. Valerio 2, 34127 Trieste, Italy\\

$\dag$ Presently at: Department of Physics, Central Michigan University, Mt. Pleasant, MI 48859, US\\

$\ddag$ Presently at: School of Physics \& Astronomy, University of Nottingham, Nottingham NG7 2RD, UK\\\\
 {*}corresponding author:
Virginia Carnevali (carne1v@cmich.edu)\\

\begin{abstract}
 Two-dimensional materials on metallic surfaces or stacked one on top of the other can form a variety of moir\'e superstructures depending on the possible
 parameter and symmetry mismatch and misorientation angle. In most cases, such as incommensurate lattices or identical lattices but with a small twist angle, the common periodicity may be very large, thus making numerical simulations prohibitive. We propose here a general procedure to determine the minimal simulation cell which approximates, within a certain tolerance and a certain size, the primitive cell of the common superlattice, given the two interfacing lattices and the relative orientation angle. As case studies to validate our procedure, we report two applications of particular interest: the case of misaligned hexagonal/hexagonal identical lattices, describing a twisted graphene bilayer or a graphene monolayer grown on Ni(111), and the case of
 hexagonal/square lattices, describing for instance a graphene monolayer grown on Ni(100) surface. The first one, which has also analytic solutions, constitutes a solid benchmark for the algorithm; the second one shows that a very nice description of the experimental observations can be obtained also using the resulting relatively small coincidence cells.

\end{abstract}
\maketitle

\section{Introduction}
A moir\'e superstructure can be defined as the interference figure due to the overlap of two or more patterns \cite{moire}. Such structures can be  commonly observed in surface science when, for example, a layer of a material is supported on a substrate
having a different lattice periodicity \cite{ZZ,Marchini,Sule,Busse,Reidy}.
The issue of stacking different lattices is involved also in
the formation of novel materials through layer-by-layer combinations of 2D systems \cite{GrR1,GrR2,FHe}, a field nowadays in rapid expansion after
the discovery of graphene \cite{GR}.
The resulting heterostructures, possibly grown on different substrates \cite{sub}, are held together by weak Van der Waals forces and can be easily manipulated at the level of each single layer \cite{exf}.
Due to their peculiar and tunable electronic properties, these heterostructures  are new promising materials for the realization of advanced and innovative electronic devices, such as field-effect transistors, light-emitting diodes, nonvolatile memory cells, and other small-sized devices where interface properties are important \cite{p-n,diode,GrT,MoS2Gr}.

 A deep knowledge and characterization of these structures is increasingly necessary both from an experimental and theoretical point of view.
Among the theoretical approaches, the first principle Density Functional Theory (DFT) is one of the most accurate, but it strongly relies on the possibility to adequately model the primitive cell of such systems, that is usually a difficult task.
Most of the state-of-the-art DFT codes for extended systems make use of periodic boundary conditions and are formulated in the reciprocal space.
Therefore, it is necessary to identify a primitive cell as close as possible to the real system structure, and this is particular challenging for heterostructures where the underlying Bravais lattices of the 2D  stacked materials are mismatched or with a small twist orientation angle \cite{Kaxiras}.
Implementing an appropriate primitive cell for DFT simulations is really important: small variations in the cell size can have a significant
impact in the final configuration and description of the simulated system \cite{ZZ}.

By looking at the recent literature, some
solutions have been proposed for a restricted subset of heterostructures, mostly regarding the hexagonal lattice interfaced with others \cite{HG}, or, even more specifically, with another hexagonal lattice \cite{HH} and in particular with the same lattice parameter \cite{Hex}. A general solution for two generic 2D Bravais lattices has been formulated in the reciprocal space \cite{Rec}, or in the real space only when experimental hints can be provided \cite{Artaud}. A few number of open-source codes
are available to build a primitive cell of a moir\'e superstructure taking into account a strain condition of one lattice with
respect to the other, searching and sorting results within given combinational spaces \cite{Stradi,CellMatch}.\

In this paper we are going to present a general deterministic algorithm that allows to identify the primitive cell of a moir\'e superstructure generated by
two generic 2D Bravais lattices, given their misorientation angle. This is done
in a subspace of cells with limited size and a certain tolerance on the coincidence condition
by solving a set of equations in real space.
The proposed model is purely geometric, and does not take into account any interaction between the two lattices.
We show that the smallest  tolerance threshold that allows solutions within a finite space
 gives a quite realistic primitive cell of the system in comparison to the experiment. A similar approach has been proposed in Ref.\cite{Koda} where it has been validated for 2D crystals and van der Waals-bonded heterostructures. We describe here in detail our method, step by step, and provide a sample code in Python that can be easily adapted and generalized.
Our method has been tested for the specific cases
of hexagonal/hexagonal misaligned identical lattices, describing a twisted graphene bilayer or a graphene monolayer grown on Ni(111), and the case of
 hexagonal/square lattices, representative of graphene grown on Ni(100) surface, finding in all cases an excellent agreement with the specific literature \cite{JEROMY,ZZ,Hex}.

\section{Results and discussion}
We consider two generic 2D Bravais lattices generated by the primitive vectors ($\mathbf{a_{o_1}}$,$\mathbf{a_{o_2}}$) and ($\mathbf{a_{s_1}}$,$\mathbf{a_{s_2}}$), where $o$ and $s$ refer to "overlayer" and "substrate" respectively, without any loss of generality.

\begin{figure}[h]
\centering
\includegraphics[scale=0.65]{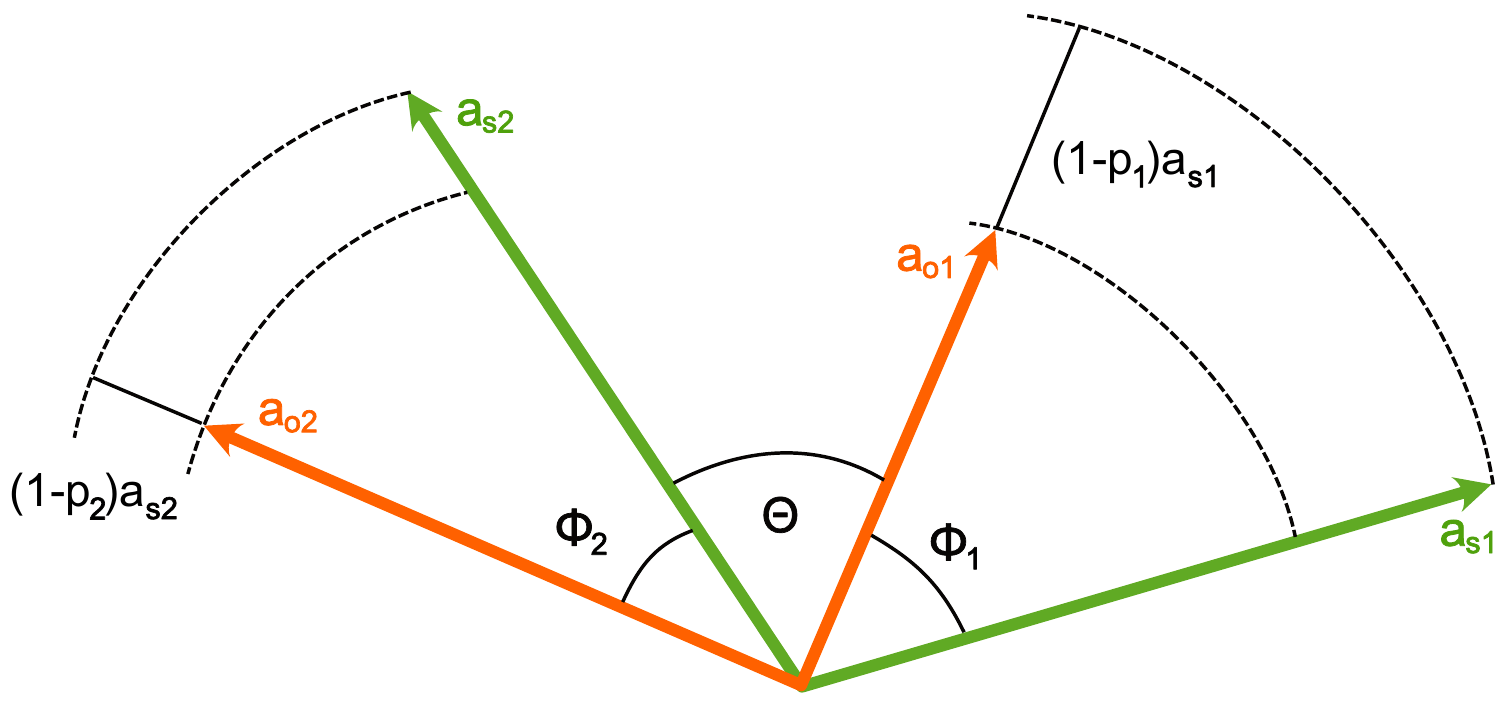}
\caption{Schematic representation of the basis vectors for two stacked generic 2D Bravais lattices, together with the geometrical parameters $\phi_1, \phi_2, \theta, p_1,\, \mbox{and} \,p_2, $ related to the mismatch conditions. }
\label{fig:1}
\end{figure}

Since we are interested in identifying common lattice vectors, if exist, it is useful to express the basis on one stacked lattice with respect to the other. For the sake of definiteness we make a specific choice: in particular, we write the primitive vectors of the overlayer with respect to those of the substrate, since this corresponds to the real situation where the overlayer adopts to the substrate, but we could do the alternative choice as well. Therefore, we can always write:\\
\begin{equation}\label{eq:stacked}
 \begin{pmatrix}
  \mathbf{a_{o_{1}}} \\
  \mathbf{a_{o_{2}}}
 \end{pmatrix}
    =\begin{pmatrix}
   a & b \\
   c & d
  \end{pmatrix}
  \begin{pmatrix}
   \mathbf{a_{s_{1}}} \\
   \mathbf{a_{s_{2}}}
  \end{pmatrix}
\end{equation}
with $(a,b,c,d)$  {\it real} numbers, in general.
The relative superposition of the two stacked lattices can also be equivalently described
using  physically meaningful parameters:
 the misorientation angle
$\phi_1 = \angle(\mathbf{a_{o_1}},\mathbf{a_{s_1}})$,
two other angles related to it and to the two sets of primitive vectors
($\phi_2=\angle(\mathbf{a_{o_2}},\mathbf{a_{s_2}})$,
$\theta=\angle(\mathbf{a_{o_1}},\mathbf{a_{s_2}}))$    and the scaling factors
of the  primitive vectors length ($p_1=|a_{o1}|/|a_{s1}|, p_2=|a_{o2}|/|a_{s2}|$) (Fig. \ref{fig:1}).
With reference to Fig. \ref{fig:1} and Eq. \ref{eq:stacked}, the scalar products $\mathbf{a_{s_\mu}}\cdot \mathbf{a_{o_\nu}} $, with $\mu,\nu=1,2$, read as:
\begin{subequations}\label{eq:abcdexact-all}
\begin{equation}\begin{aligned}\label{eq:abcdexact-a}
\mathbf{a_{s_1}}\cdot \mathbf{a_{o_1}}&= a |\mathbf{a_{s_1}}|^2 + b |\mathbf{a_{s_1}}||\mathbf{a_{s_2}}|\cos(\phi_1 + \theta)\\
&= |\mathbf{a_{s_1}}||\mathbf{a_{o_1}}|\cos(\phi_1)
\end{aligned}\end{equation}
\begin{equation}\begin{aligned}\label{eq:abcdexact-b}
\mathbf{a_{s_2}}\cdot \mathbf{a_{o_1}}&= b |\mathbf{a_{s_2}}|^2 + a |\mathbf{a_{s_1}}||\mathbf{a_{s_2}}|\cos(\phi_1 + \theta)\\
&= |\mathbf{a_{s_2}}||\mathbf{a_{o_1}}|\cos(\theta)
\end{aligned}\end{equation}
\begin{equation}\begin{aligned}\label{eq:abcdexact-c}
\mathbf{a_{s_1}}\cdot \mathbf{a_{o_2}}&= c |\mathbf{a_{s_1}}|^2 + d |\mathbf{a_{s_1}}||\mathbf{a_{s_2}}|\cos(\phi_1 + \theta)\\
&= |\mathbf{a_{s_1}}||\mathbf{a_{o_2}}|\cos(\phi_1 + \phi_2 +\theta)
\end{aligned}\end{equation}
\begin{equation}\begin{aligned}\label{eq:abcdexact-d}
\mathbf{a_{s_2}}\cdot \mathbf{a_{o_2}}&= d |\mathbf{a_{s_2}}|^2 + c |\mathbf{a_{s_1}}||\mathbf{a_{s_2}}|\cos(\phi_1 + \theta)\\
&= |\mathbf{a_{s_2}}||\mathbf{a_{o_2}}|\cos(\phi_2)
\end{aligned}\end{equation}
\end{subequations}
The latter constitutes a linear system of four equations in the four unknowns $(a,b,c,d)$ and one can easily show that the coefficient matrix is full rank (because the angle $\phi_1 + \theta$ cannot be a multiple of $\pi$).
In other words, given a pair of stacked 2D Bravais lattices with a certain misorientation condition and thus knowing the quantities $\phi_1, \phi_2, \theta,$ $ p_1,\, \mbox{and}\,  p_2$,
the system of  Eq. \ref{eq:abcdexact-all} has  always {\it real} solutions for  the parameters $(a, b, c, d)$, that we refer to as  $(a,b,c,d)^{exact}$.

\indent Once defined the relationship between the two 2D Bravais lattices, the existence of a common  moir\'e superstructure has to be discussed. To do that, we consider now the
infinite sets of vectors  $\{\mathbf{a_{om_{\nu}}}\}$ and  $\{\mathbf{a_{sm_{\mu}}}\}$ $(\nu,\mu=1,2)$ generated by the primitive vectors  of the stacked lattices:
\begin{equation}\begin{gathered}
\label{eq:ijklmnqr}
 \begin{pmatrix}
 \mathbf{a_{om_{1}}} \\
  \mathbf{a_{om_{2}}}
 \end{pmatrix} =
 \mbox{M}_{o}
 		\begin{pmatrix}
               \mathbf{a_{o_{1}}} \\
               \mathbf{a_{o_{2}}}
              \end{pmatrix} =
              \begin{pmatrix}
                  i & j \\
                  k & l
                 \end{pmatrix}
                  		\begin{pmatrix}
               \mathbf{a_{o_{1}}} \\
               \mathbf{a_{o_{2}}}
              \end{pmatrix}\\
                   \quad \mbox{and} \quad \\
  \begin{pmatrix}
 \mathbf{a_{sm_{1}}} \\
  \mathbf{a_{sm_{2}}}
 \end{pmatrix} =\mbox{M}_{s} \begin{pmatrix}
               \mathbf{a_{s_{1}}} \\
               \mathbf{a_{s_{2}}}
              \end{pmatrix} =
              \begin{pmatrix}
                m & n \\
                q & r
               \end{pmatrix}
              \begin{pmatrix}
               \mathbf{a_{s_{1}}} \\
               \mathbf{a_{s_{2}}}
              \end{pmatrix}
\end{gathered}\end{equation}
with $(i,j,k,l)$ and $(m,n,q,r)$ integers.

~\\ In case of commensurability, vectors common to the two sets
 and corresponding common multiple cells can be found.
 Labelling the moir\'e  primitive vectors   with ($\mathbf{a_{m_1}}$, $\mathbf{a_{m_2}}$),
for proper values of the matrices $M_o$ and $M_s$, the commensurability condition reads
\cite{Artaud}:

 \begin{equation}
 \begin{gathered}
 \begin{aligned}
\label{eq:commensurability}
 \begin{pmatrix}
 \mathbf{a_{m_{1}}} \\
  \mathbf{a_{m_{2}}}
 \end{pmatrix} &=
 \mbox{M}_{o}\begin{pmatrix}
               \mathbf{a_{o_{1}}} \\
               \mathbf{a_{o_{2}}}
              \end{pmatrix} =
               \begin{pmatrix}
                  i & j \\
                  k & l
                 \end{pmatrix} \begin{pmatrix}
               \mathbf{a_{o_{1}}} \\
               \mathbf{a_{o_{2}}}
              \end{pmatrix}\\
   &=\mbox{M}_{s}\begin{pmatrix}
               \mathbf{a_{s_{1}}} \\
               \mathbf{a_{s_{2}}}
              \end{pmatrix} =
                  \begin{pmatrix}
                  m & n \\
                  q & r
                 \end{pmatrix} \begin{pmatrix}
               \mathbf{a_{s_{1}}} \\
               \mathbf{a_{s_{2}}}
              \end{pmatrix}
\end{aligned}
\end{gathered}
\end{equation}

Since the two primitive vectors ($\mathbf{a_{m_1}}$, $\mathbf{a_{m_2}}$) cannot be parallel, both matrices $M_o$ and $M_s$ have to be invertible ($il-jk \neq 0$ and $mr-nq \neq 0$).
Eq. \ref{eq:commensurability} implies a specific relationship between the two stacked lattices, that, together with the generic  Eq. \ref{eq:stacked}, gives:
\begin{subequations}\label{eq:abcd-rational-all}
\begin{equation}\begin{aligned}\label{eq:abcd-rational-a}
 \begin{pmatrix}
  \mathbf{a_{o_{1}}} \\
  \mathbf{a_{o_{2}}}
 \end{pmatrix}=
  \mbox{M}_{o}^{-1}\mbox{M}_{s} \begin{pmatrix}
               \mathbf{a_{s_{1}}} \\
               \mathbf{a_{s_{2}}}
              \end{pmatrix}
  =\begin{pmatrix}
   a & b \\
   c & d
  \end{pmatrix}
  \begin{pmatrix}
   \mathbf{a_{s_{1}}} \\
   \mathbf{a_{s_{2}}}
  \end{pmatrix}\\
  \end{aligned}\end{equation}
\begin{equation}\begin{aligned}\label{eq:abcd-rational-b}
\centering
 \mbox{where}\qquad &\\
 a &=\frac{lm-jq}{il-jk},\quad b =\frac{ln-jr}{il-jk},\\
 c &=\frac{-km+iq}{il-jk},\quad d =\frac{-kn+ir}{il-jk}.
\end{aligned}\end{equation}
\end{subequations}
In conclusion, Eq. \ref{eq:abcd-rational-b} shows that the four parameters $(a,b,c,d)$, that are {\it real} numbers in general in Eqs. \ref{eq:stacked}-\ref{eq:abcdexact-all},  are {\it rational} numbers when the commensurability condition (Eq. \ref{eq:commensurability}) is satisfied.

~\\
We want to face now the general problem of finding an approximate coincidence lattice
even when the commensurability condition is not perfectly satisfied.
To do that,  instead of starting from Eqs. \ref{eq:stacked} and \ref{eq:abcdexact-all} expressing the exact relationship between the two stacked lattices, we start from the commensurability condition, expressed by  Eqs. \ref{eq:commensurability} and \ref{eq:abcd-rational-all}, and search for
a set of parameters $(a, b, c, d)$ that are {\it rational} and are {\it approximate} solutions of Eq. \ref{eq:abcdexact-all},
i.e, deviate from the exact real values
$(a,b,c,d)^{exact}$ within a certain \textcolor{black}{tolerance $t$. More precisely, for any rational 4-tuple $(a, b, c, d)$ we define a distance $\delta$ as follows $$\delta \equiv \max\{|a-a^{exact}|, |b-b^{exact}|, |c-c^{exact}|, |d-d^{exact}|\} , $$ and require that $\delta < t$. }
From a physical point of view, such tolerance
can be interpreted as a strain on the overlayer
to force the matching with the substrate
over a certain number of periods indicated by  $M_o$ and $M_s$,
so that the equality between the first and the second row of Eq.  \ref{eq:commensurability} holds. Therefore, the threshold on  $(a,b,c,d)$ is a threshold on
commensurability, and
 acceptable values have thus to be set
on physical grounds, considering the elasticity of the overlayer.

As it can be easily seen from Eq. \ref{eq:abcd-rational-b},  there is an infinite number of combinations of the eight integers ($i,j,k,l,m,n,q,r$) that can give the same value of parameters $(a, b, c, d)$ equal or close to
$(a,b,c,d)^{exact}$. \textcolor{black}{In order to keep the computation finite, we constrain each of the eight integers to vary in the interval $[-R,R]$. The consequences of this restriction and the rationale behind a good choice of the parameter $R$ are discussed in the next section. Given a certain value for the parameters $t$ and $R$, we call $I^{t,R} \subset \mathbb{Z}^8$ the set of integers that ensure $\delta <t$. In the following discussion, we always assume that $t$ and $R$ are fixed a priori and, therefore, we commonly refer to $I^{t,R}$ as $I$ in order to ease the notation. }
For future convenience, we are also interested in considering separately the first four numbers and the last four in each element of $I$, therefore we define the two sets $I_s = \{ (i,j,k,l) | (i,j,k,l,m,n,q,r) \in I \}$ and $I_o = \{ (m,n,q,r) | (i,j,k,l,m,n,q,r) \in I \}$.
Among such combinations, we are looking for those solutions that give the smallest possible cells. The area of a cell can be computed from the lattice vectors as follows
\begin{align}
\label{eq:Acell}
& A^s_{cell} = |\mathbf{a_{sm_1}} \times \mathbf{a_{sm_2}}| = |m\mathbf{a_{s_1}} \times r\mathbf{a_{s_2}} + q\mathbf{a_{s_2}} \times n\mathbf{a_{s_1}}|=| mr - qn | ~| \mathbf{a_{s_1}} \times \mathbf{a_{s_2}} |, \nonumber \\
&A^o_{cell} = |\mathbf{a_{om_1}} \times \mathbf{a_{om_2}}| = |i\mathbf{a_{o_1}} \times l\mathbf{a_{o_2}} + j\mathbf{a_{o_2}} \times k\mathbf{a_{o_1}}|=| il - jk |~ | \mathbf{a_{o_1}} \times \mathbf{a_{o_2}}|  ,
\end{align}
where the vectors $\mathbf{a_{sm_1}}, \mathbf{a_{sm_2}}, \mathbf{a_{om_1}}, \mathbf{a_{om_1}}$ have been defined in Eq. \ref{eq:ijklmnqr}.
In case of perfect commensurability, as described in Eq. \ref{eq:commensurability}, there exists a choice of those vectors that allows to define a moir\'e supercell and its area is consistently given by $A_{cell} = |\mathbf{a_{m_1}} \times \mathbf{a_{m_2}}| = A^o_{cell} = A^s_{cell}$. In the more general setting we are interested in, one can nevertheless compute separately $A^o_{cell}$ and $A^s_{cell}$ given an element of $I$. Those areas will be different but sufficiently close if the commensurability threshold is small. One can then minimize $A^o_{cell}$ over $I_o$ and $A^s_{cell}$ over $I_s$ in order to find the primitive cells
\begin{align}
\label{eq:Amin}
 A^o_{min} &= \min\limits_{\substack{(i,j,k,l)} \in I_o}
| il - jk |~ | \mathbf{a_{o_1}} \times \mathbf{a_{o_2}}| , \nonumber\\
 A^s_{min} &= \min\limits_{\substack{(m,n,q,r)} \in I_s}
| mr - qn | ~| \mathbf{a_{s_1}} \times \mathbf{a_{s_2}} |.
\end{align}
Again, for a resonable threshold, the two quantities $A^o_{min}$ and $A^s_{min}$ will be approximately equal and the two minimizers will be consistent, i.e. $(i,j,k,l)_{min}$ and $(m,n,q,r)_{min}$ will be such that $(i,j,k,l,m,n,q,r)_{min} \in I$.
One the two cells is then conventionally chosen as the approximate moir\'e supercell.

The system of equations Eq.~\ref{eq:abcd-rational-all}, dubbed "moir\'e superlattice relations" or "commensurability conditions", relating the superlattice parameters with the geometry of the system, together with Eq.~\ref{eq:abcdexact-all}  and the minimization condition of Eq.~\ref{eq:Amin} constitutes the core of our systematic procedure to obtain the smallest moir\'e primitive cell.

Different ($i, j, k, l, m, n, q, r$) solutions of Eq.~\ref{eq:Amin} identify moir\'e primitive cells with different shape and orientation, but this corresponds to the fact that the choice of the primitive vectors for a Bravais lattice is not unique.

We report in the next section the details of the procedure.

\begin{figure}[ht]
 \centering
\includegraphics[scale=0.4]{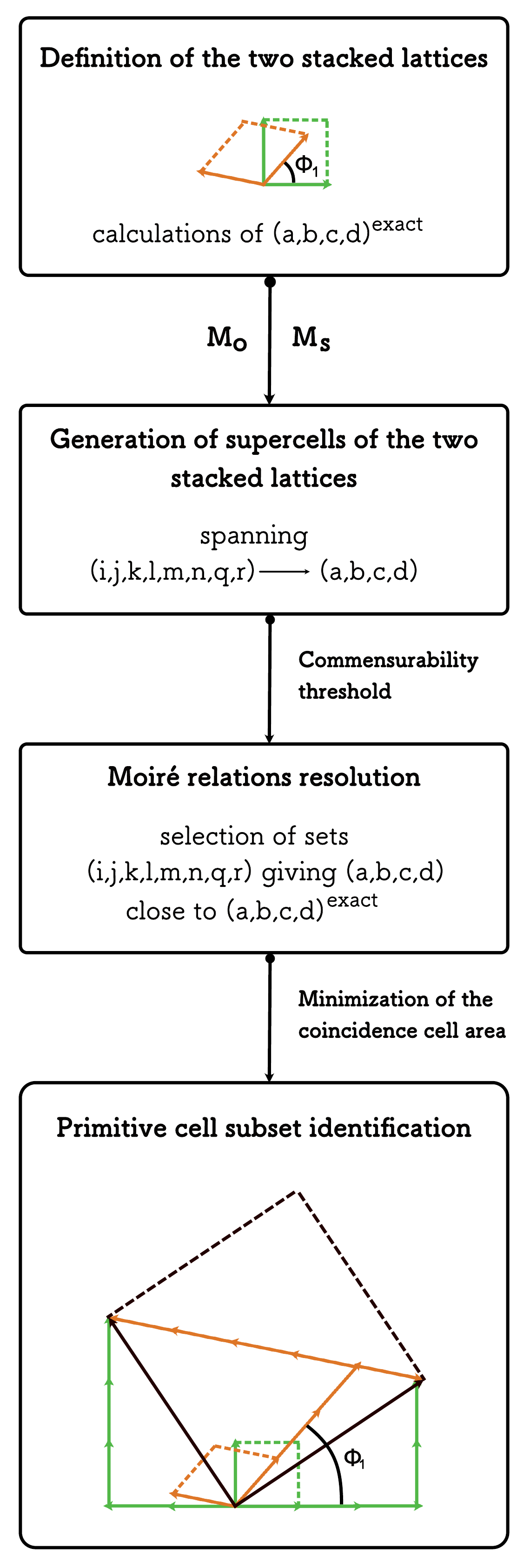} 
\caption{Flowchart  of the algorithm for finding the primitive cell of a moir\'e superstructure.
(i): the two
lattices together with their misorientation condition are defined, and
the real parameters $(a,b,c,d)^{exact}$ giving exactly one layer with respect to the other are calculated.
(ii): through the matrix $M_{o}$ and $M_{s}$ (indexes $(i, j, k, l, m, n, q, r)$ varying within a certain range)
possible supercells of the two stacked lattices are generated. The parameters  $(a,b,c,d)$ are calculated.
(iii): considering a commensurability threshold, a subset of
possible primitive cells is identified by comparing  $(a,b,c,d)$  with $(a,b,c,d)^{exact}$; the corresponding
sets of indexes $(i, j, k, l, m, n, q, r)$ is considered to describe a common coincidence cell.
(iv): the minimization on the area of the cells determined by the sets $(i, j, k, l, m, n, q, r)$ selected in (iii)
provides the smallest primitive cells for the moir\'e superstructure compatible with the fixed commensurability tolerance and size.}
\label{fig:2}
\end{figure}

\subsection{Algorithm details}
The procedure previously mentioned can be easily implemented in a code, whose flowchart can be summarized in four main points, as schematically reported  in Fig. \ref{fig:2}.

\begin{enumerate}
   \item[(i)] \textit{Definition of the primitive cells of the two stacked lattices and their relative orientation.}\\
 First of all, the code reads from input the two 2D Bravais lattices, i.e. their basis vectors, and the misorientation angle. Through Eq. \ref{eq:abcdexact-all}, the code computes the exact value $(a,b,c,d)^{exact}$ of the parameters $(a,b,c,d)$.
   \item[(ii)] \textit{Generation of supercells for the two stacked lattices.}\\
 The code considers all the 8-tuples of integers $(i, j, k, l, m, n, q, r)$
 spanning over the range $[-R,R]$ that has to be fixed;
this corresponds to consider supercells of the substrate  (through $(i, j, k, l)$) and of the overlayer
  (through $(m, n, q, r)$).
 For each 8-tuple, the corresponding rational parameters $(a,b,c,d)$ are calculated according to Eq.~\ref{eq:abcd-rational-b}. This  calculation is completely independent of the specific problem considered and clever methods should be implemented to perform it once and for all.
   \item[(iii)] \textit{Moir\'e relations solution.}\\
The parameters $(a,b,c,d)$ obtained in step (ii) for each 8-tuple are used to check the condition $\delta < t$, for a chosen $t$. If the inequality is satisfied, the corresponding 8-tuple is included in the solution ensemble $I$.

   \item[(iv)] \textit{Primitive cells subset identification.}\\
   The last step is to identify among the elements of $I$ those associated with the smallest cells $A^o_{min}$ and $A^s_{min}$, with $A^o_{min} \sim A^s_{min}$, defined through the minimization procedure in Eq. \ref{eq:Amin}. If the threshold $t$ is sufficiently small and the parameter $R$ is sufficiently large, the two resulting cells are very similar and one of the two can be conveniently chosen as the moir\'e supercell.
Following the physical picture of substrate and overlayer, the reasonable choice
is the one defined by the substrate.
 \end{enumerate}

 \textcolor{black}{ The algorithm we synthetically presented can be applied to any choice of the substrate and overlayer, and it only requires as a \textit{physical input} the geometry of the problem (the primitive vectors of the two lattices and their relative orientation). We notice, however, that a delicate point is the choice of the two \textit{computational parameters}, namely the integer spanning range $R$ and the commensurability threshold $t$. These two parameters describe a tradeoff between accuracy of the result and employed computational resources. \\
Indeed, a finite range $R$ is needed in order to keep the computation finite and, practically, one would like to have it small because the number of configurations to be checked grows as $N^8$, with $N=2R+1$. Also, one is not interested in a cell that is possibly too large to be used for ab-initio calculations, and the highest possible area in a fixed range is proportional to $2R^2$ (see eq. \ref{eq:Acell}). However, a large $R$ is not automatically associated with a large cell area, because for instance $|il-jk|=1$ for $i=j=1$ and $l=R, k=R-1$. Moreover, the wider is $R$, the higher is the possibility of finding sets that describe a primitive cell close to the perfect coincidence moir\'e superstructure. \\
 The closeness is controlled by the threshold $t$. If $t$ is small, the accepted cells describe a very accurate moiré pattern, but if it is too small, no solutions are reasonably expected when the geometry does not allow for perfect commensurability (see our case study n.2).  \\
The discussion above is to clarify that a sensible and balanced choice of the \textit{computational parameters} is highly nontrivial and the problem is general, namely it is not specific to our proposed algorithm. In this respect, our proposal has the advantage of being suitable for parallel computing, by fixing the amount of computations to be performed once and for all.
In any case, despite being case dependent, in order to start the computation we expect a good tradeoff to be $R=10$, $t=0.05$. This is because it excludes from the beginning too large areas that are unwanted in any case and nevertheless it reasonably provides many possible solutions.
 }


\subsection{Case study n. 1: hexagonal/hexagonal misaligned identical lattices}
For the sake of definiteness we refer here to the case of twisted graphene bilayer, but the case
of rotated graphene over Ni(111) could be equivalently considered,
if the very small mismatch of the lattice parameters (smaller than 2\%) is neglected.
The twisted graphene bilayer presents different moir\'e superstructures according to the relative rotational angle between the two lattices. We consider the angle between the primitive vectors of each lattice,
$\angle(\mathbf{a_{s_1}},\mathbf{a_{s_2}})$ and $\angle(\mathbf{a_{o_1}},\mathbf{a_{o_2}})$,
equal to $120^\circ$.
 Dealing with two identical hexagonal lattices we have $|\mathbf{a_{s_1}}|=|\mathbf{a_{s_2}}|=|\mathbf{a_{o_1}}|=|\mathbf{a_{o_2}}|=L$, $\phi_2=\phi_1$ and $\phi_1+\theta= 120^\circ$. As a result, Eq. \ref{eq:abcdexact-all}  give:
\begin{subequations}
\label{eq:Aminall}
\begin{equation}\label{eq:Aminaa}
a= \cos(\phi_1) + \frac{\sqrt{3}}{3}\sin(\phi_1)
\end{equation}
\begin{equation}\label{eq:Aminb}
b= \frac{2\sqrt{3}}{3}\sin(\phi_1)
\end{equation}
\begin{equation}\label{eq:Aminc}
c= -\frac{2\sqrt{3}}{3}\sin(\phi_1)
\end{equation}
\begin{equation}\label{eq:Amind}
d= \cos(\phi_1)-\frac{\sqrt{3}}{3}\sin(\phi_1)
\end{equation}
\end{subequations}
and the minimization of the possible moir\`e cell (Eq. \ref{eq:Amin}) reads:

\begin{align}
\label{eq:Aminter}
 A^o_{min} &= \frac{L^2\sqrt{3}}{2} \min\limits_{\substack{(i,j,k,l)} \in I_o}
| il - jk | , \nonumber\\
A^s_{min} &= \frac{L^2\sqrt{3}}{2} \min\limits_{\substack{(m,n,q,r)} \in I_s}
| mr - qn | .
\end{align}

\begin{figure}[h]
\centering
\includegraphics[scale=0.52]{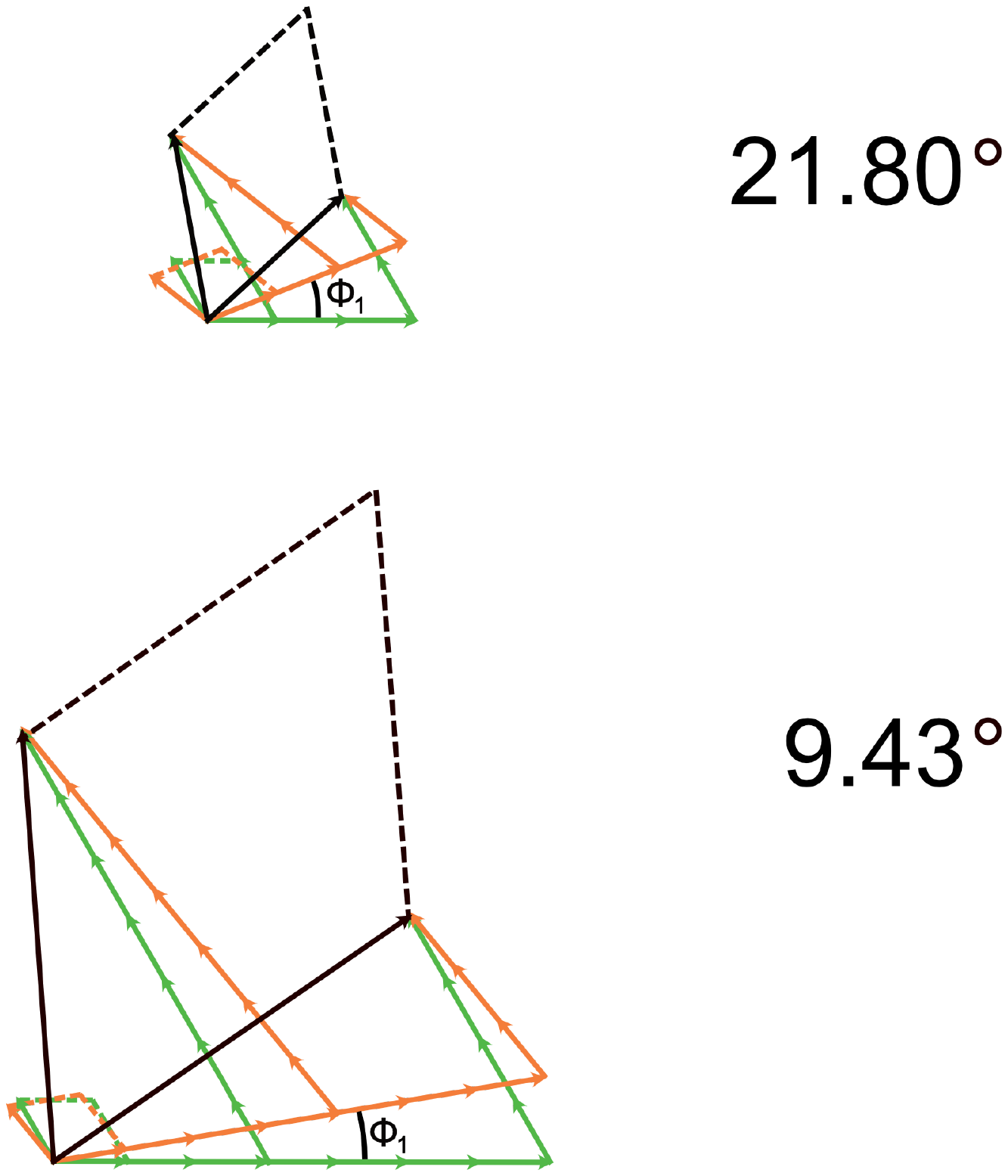}
\caption{Schematic representation of the lattice stacking in the twisted graphene bilayer for two different misorientation angles corresponding to commensurate rotations, with the indication of the primitive cells of the two interfacing lattices,  the moir\'e pattern, and its relationship with them. }
\label{fig:HexHex_cells}
\end{figure}

Following Moon \textit{et al.} \cite{Hex}, an analytic solution is possible for this system. In particular, Eq. 2 of Ref. \cite{Hex} identifies a set of rotational angles $\phi_1$
ensuring a perfect commensurability  between the two graphene lattices. According to our model, this is equivalent to find a set of primitive cells for the moir\'e superstructure where the threshold is set to 0. We have solved Eq. \ref{eq:Aminall} for two selected commensurate rotation angles, $\phi_1=21.8^\circ$
and $9.43^\circ$
with a threshold of 10$^{-7}$, equal to the floating point standard single precision,
obtaining the same primitive cells found by Moon \textit{et al.} \cite{Hex}. The results are reported in Tab. \ref{TabHex} and shown in Fig. \ref{fig:HexHex_cells}.

\begin{table}
\centering
\renewcommand\arraystretch{1.3}
\begin{tabular}{c|cccc}
  \cline{1-4}
  $\phi_1$ & $21.8^\circ$ & $9.43^\circ$&  \\ \cline{1-3}
  $M_{o}$     &    $\begin{pmatrix} 3 & 1
\\ 2 & 3\end{pmatrix}$          &   $\begin{pmatrix} 7 & 3
\\ 4 & 7 \end{pmatrix}$      &  \\
  $M_s$        &    $\begin{pmatrix} 3 & 2
\\ 1 & 3 \end{pmatrix}$        &      $\begin{pmatrix} 7 & 4
\\ 3 & 7 \end{pmatrix}$   &  \\
  $N_{s}$        &     7    &     39          & \\ \cline{1-3}
\end{tabular}
\caption{Summary of the results obtained for the coincidence lattice of the twisted graphene bilayer for different misorientation angle $\phi_1$ with a coincidence tolerance threshold $t=10^{-7}$. The matrices $M_{o}$. $M_s$ giving the unit cell of the moir\'e superstructures and the number of elementary primitive cells $N_{s}, N_{o}$ of each graphene layer (substrate and overlayer) contained in the moir\'e cell are reported. In this particular case, $N_{s}=N_{o}$ and
$M_{o}^T=M_s$ and viceversa,
since substrate and overlayer here are exactly the same lattice.}
\label{TabHex}
\end{table}

\subsection{Case study n. 2: hexagonal/square}
Also the graphene/Ni(100) system, prototype of hexagonal/square lattice stacking, shows different moir\'e superstructures tuned by the misorientation between the different interfacing lattices.
As opposed to the case of the twisted graphene bilayer, graphene/Ni(100) does not present any rotational angle that allows a perfect commensurability between the interfacing lattices, as explained in more detail below.

The basis vectors of the graphene hexagonal lattice have the same modulus and the same applies to the Ni(100) square lattice. The angles between the primitive vectors of each lattice are
$\angle(\mathbf{a_{s_1}},\mathbf{a_{s_2}})=90^\circ$ and $\angle(\mathbf{a_{o_1}},\mathbf{a_{o_2}})=120^\circ$, so that one also has $\theta= 90^\circ - \phi_1$ and $\phi_2=\phi_1+30^\circ$.
Therefore, considering $|\mathbf{a_{s_1}}|=|\mathbf{a_{s_2}}|=L$ and $|\mathbf{a_{o_1}}|=|\mathbf{a_{o_2}}|=p L$,
 the Eq. \ref{eq:abcdexact-all} read as follows:
\begin{subequations}
\label{eq:7all}
\begin{equation}\label{eq:7a}
a= p \cos(\phi_1) ,
\end{equation}
\begin{equation}\label{eq:7b}
b= p \sin(\phi_1) ,
\end{equation}
\begin{equation}\label{eq:7c}
c= -p \left(\frac{\sqrt{3}}{2}\sin(\phi_1)+ \frac{1}{2}\cos(\phi_1) \right) ,
\end{equation}
\begin{equation}\label{eq:7d}
d= p \left( \frac{\sqrt{3}}{2}\cos(\phi_1)-\frac{1}{2}\sin(\phi_1) \right) ,
\end{equation}
\begin{equation}
\label{eq:7e}
 A^o_{min} = L^2 p^2\frac{\sqrt{3}}{2} \min\limits_{\substack{i,j,k,l} \in I_o}
|il-jk|,
\end{equation}
\begin{equation}
\label{eq:7f}
A^s_{min} = L^2 \min\limits_{\substack{m,n,q,r} \in I_s}
  |mr-qn|.
\end{equation}
\end{subequations}

Given the lattice parameters of graphene and Ni(100), 2.46 \AA{} and 2.49 \AA{} respectively, the scaling factor of the system turns out to be $p = 0.988$. From the equations above, one can prove by contradiction that, contrary to the previous example, perfect commensurability is impossible in this case. Indeed, combining for instance Eqs. \ref{eq:7a}, \ref{eq:7b} and \ref{eq:7d} one finds $2d + b = \sqrt{3}a$. Assuming $a,b,c,d$ to be rational, and $a\neq 0$, one arrives at a contradiction, because the left-hand side of the equality $2d+b$ is rational, while the right-hand side is not, as the product of a nonzero rational number $a$ and an irrational one $\sqrt{3}$.
If instead $a=0$, one has $2c= - b \sqrt{3}$. Since $a$ and $b$ cannot vanish simultaneously, a contradiction arises also in this case. Therefore $a,b,c,d$ cannot be simultaneously rational and the commensurability conditions, Eq. \ref{eq:abcd-rational-b}, are never satisfied.

Note that for this heterostack the configuration with a rotational angle of $\phi_1$ corresponds
to that with $60^\circ - \phi_1$ up to a redefinition of the primitive vectors. We use this fact in order to compare the theoretical results with the experimental observations.
In particular, we compute explicitly the moir\'e relations for
three selected orientations, namely
$\phi_1=45.26^\circ$
(14.74$^\circ$ in Ref. \cite{JEROMY}), $48.7^\circ$ (11.3$^\circ$ in Ref. \cite{ZZ}), and $54.71^\circ$
(5.29$^\circ$ in Ref. \cite{JEROMY}).

\begin{figure}[h]
\centering
\includegraphics[scale=0.52]{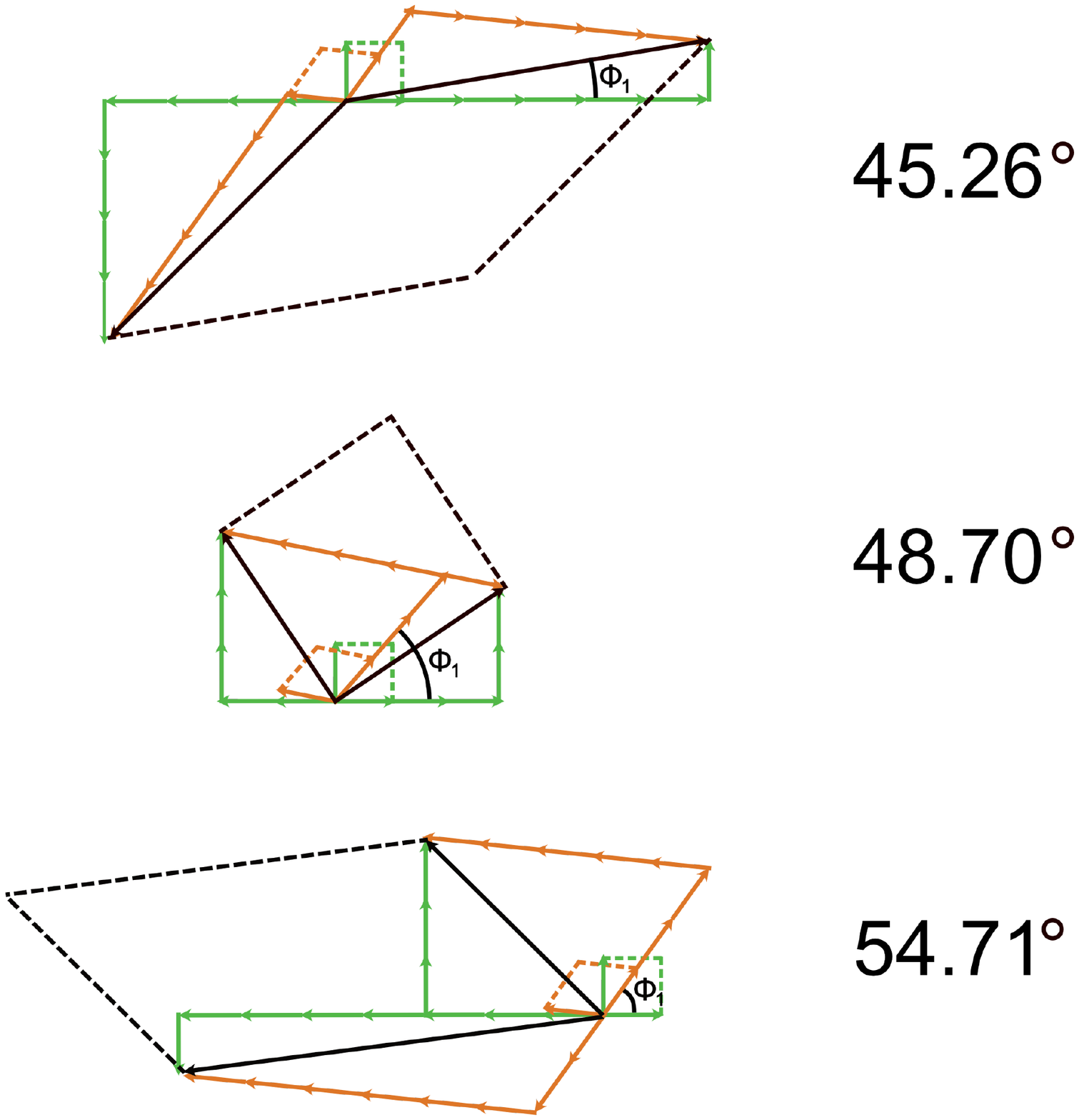}
\caption{Schematic representation of the lattice stacking in graphene/Ni(100) for three different misorientation angles corresponding to commensurate rotations, with the indication of the primitive cells of the two interfacing lattices, the moir\'e pattern, and  its relationship with them. }
\label{fig:HexSq_cells}
\end{figure}

For each angle $\phi_1$ we report in Tab. \ref{TabAng} and show in Fig. \ref{fig:HexSq_cells}
the smallest primitive cell  obtained with the
minimization procedure of our algorithm using  a threshold $t=0.04$ on the $(a,b,c,d)$ parameters.
In Tab. \ref{TabAng} we report the matrices $M_{o}$ and $M_s$, the number of unitary elementary cells of the
substrate and of the overlayer, and, considering that in each unit cell of graphene there are 2 carbon atoms,
 also the ratio between the number of carbon atoms in the overlayer and the surface nickel atoms of the substrate, $C/Ni=2N_{o}/N_{s}$, to ease the comparison with Ref. \cite{JEROMY}.

For $\phi_1=48.7^\circ$ the moir\'e superstructure is a square network and the smallest cell found by our algorithm describes very well the experimental images \cite{ZZ}.
For $\phi_1=45.26^\circ$ and $\phi_1=54.71^\circ$, we can compare our results with the cells proposed in \cite{JEROMY}. The latters are much wider than ours, as we can see from
 the number of unitary elementary cells of the
substrate and of the overlayer in the moir\'e cells proposed there, taken from Tab. 1 of \cite{JEROMY} and reported in
the last lines of our Tab. \ref{TabAng}. However, we can appreciate that the ratio $C/Ni$ from our small cells is very similar to that corresponding to the large moir\'e cells of \cite{JEROMY}.

We comment that the tolerability threshold of $0.04$ is acceptable for this system.
From Tab. \ref{TabAng} and Eq. \ref{eq:7e} and \ref{eq:7f}
we can calculate $A^o_{min}/L^2$  obtaining 19.44, 12.68 and 23,67 for
$\phi_1=45.26^\circ , 48.7^\circ$ and $54.71^\circ$, respectively, to be compared with
$A^s_{min}/L^2$ that is 20, 13 and 24 for the three angles.
The relative difference between $A^o_{min}/L^2$  and $A^s_{min}/L^2$,
a quantity that somehow quantify the strain
on the overlayer for forcing the matching with the substrate over an approximately common cell,
is therefore only 2.8\%, 2.5\%, and 1.4\% for the three angles,
 compatible with the elasticity properties of graphene \cite{elGr}.

We remind that besides the smallest moir\'e cells for a given tolerance,
several other cells are given by our method.
In particular, with reference to this example, also those proposed in  \cite{JEROMY} are found.
However, in most cases, even the smallest cells  properly identified by our algorithm can describe with good accuracy the relevant properties of the physical systems,
as shown for instance in \cite{ZZ}.

\begin{table}
\centering
\renewcommand\arraystretch{1.3}
\begin{tabular}{c|cccc}
  \cline{1-4}
  $\phi_1$ & $45.26^\circ$ & $48.7^\circ$ & $54.71^\circ$ &  \\ \cline{1-4}
  $M_{o}$     &    $\begin{pmatrix} 2 & -5
\\ -5 & 1 \end{pmatrix}$          &   $\begin{pmatrix} 3 & -1
\\ 3 & 4 \end{pmatrix}$    &       $\begin{pmatrix} -2 & 6
\\ 3 & 5 \end{pmatrix}$       &  \\
  $M_s$        &   $\begin{pmatrix} 6 & 1
\\ -4 & -4\end{pmatrix}$        &      $\begin{pmatrix} 3 & 2
\\ -2 & 3 \end{pmatrix}$     &      $\begin{pmatrix} -7 & -1
\\ -3 & 3 \end{pmatrix}$       &  \\
     $N_{o}$        &     23    &     15         &       28       & \\
      $N_{s}$        &     20    &     13         &       24       & \\
      $C/Ni$ &     2.30    &    2.31         &       2.33       &\\
\cline{1-4}
$N^\ast_{o}$   &  59    &     -         &       55     & \\
$N^\ast_{s}$   &     51    &     -         &       47      & \\
$C/Ni^\ast$ &     2.34  &    -         &       2.34    &\\ \cline{1-4}
\end{tabular}
\caption{
Summary of the results obtained for the coincidence lattice of graphene/Ni(100) for different misorientation angle $\phi_1$. The Table shows: the matrices $M_{o}$, $M_s$ corresponding to  the smallest moir\'e cells
corresponding to a coincidence tolerance threshold $t=0.04$;
the number of the elementary cells of the nickel substrate ($N_{s}$) and of
the graphene overlayer ($N_{o}$) per moir\`e cell, as well as
the ratio between the number of carbon atoms in the overlayer and the surface nickel atoms of the substrate, $C/Ni$.
The quantities indicated with the asterisk in the last rows refer to the cells proposed in \cite{JEROMY}.}

\label{TabAng}
\end{table}

\section{Conclusions}
We have developed a systematic procedure for determining  primitive cells of a moir\'e superstructure generated by two generic 2D Bravais lattices with a given orientation angle, working in real space. The purpose was to identify a coincidence cell of reasonable size to be used for instance as input of ab-initio simulations,  in some cases at the price of a certain tolerance for the coincidence conditions between the two interfacing lattices and the moir\`e. In order to validate our procedure, we have investigated two case studies already reported in the literature: the twisted graphene bilayer and the graphene/Ni(100) system, representative of identical hexagonal/hexagonal and hexagonal/square lattice stacking, respectively. Different rotational angles have been investigated for both systems.
 In particular, in the former case we have chosen rotational angles that allow a perfect commensurability between the two lattices and checked that our procedure gives the correct result, pushing to zero the tolerance on the commensurability condition. The second case, graphene/Ni(100), is representative of a condition of only approximate commensurability. In this case, we opted for rotational angles
 useful for a comparison with experimental observations.
 We showed that in absence of perfect commensurability conditions, the choice for the best simulation cell  is not always unique and straightforward, but the subset of the possible simulation cells, resulting from our procedure with reasonable tolerance threshold and size range, provides satisfactory models.
 The final choice will be guided by the precision required in the description of the system and/or the available computational resources.

\section*{Acknowledgments}
We acknowledge financial support from the University of Trieste (program ``Finanziamento di Ateneo per progetti di ricerca scientifica FRA 2018''). This work has been also supported by the project  ``FERMAT - Fast ElectRon dynamics in novel hybrid organic-2D MATerials'' funded by MIUR - Progetti di Ricerca di Rilevante Interesse Nazionale (PRIN), Bando 2017 - grant 2017KFY7XF.
Computational resources have been obtained from CINECA through the ISCRA initiative and the agreement with the University of Trieste.
We thank Z. Zou and C. Africh for useful discussions.

\end{document}